# Fractional abundance and the ecology of community structure

Colleen K. Kelly[1], Michael G. Bowler[2], Jeffrey B. Joy[3] & John N. Williams[5]

[1]*Department of Zoology, South Parks Road, Oxford OX1 3PS, UK;* [2]*Department of Physics, Keble Road, Oxford OX1 3RH, UK;* [3]*Department of Biological Sciences, Simon Fraser University, Burnaby, British Columbia, Canada, V5A 1S6;* [4]*Department of Land, Air, and Water Resources University of California, One Shields Avenue, Davis, CA 95616-8627*

**The ecological principle of limiting similarity dictates that species similar in resource requirements will compete, with the superior eventually excluding the inferior competitor from the community[1-4]. The observation that nonetheless apparently similar species comprise a significant proportion of the diversity in any given community has led to suggestions that competition may not in fact be an important regulator of community structure and assembly[5,6]. Here we apply a recently introduced metric of species interaction, fractional (relative) abundance[7,8], to tree species of the tropical wet forest of Barro Colorado Island, Panamá, the particular community that inspired the original model of non-niche or 'neutral' community dynamics[9]. We show a distribution of fractional abundances between pairs of most closely related congeneric tree species differing from that expected of competitive exclusion, but also inconsistent with expectations of simple similarity, whether such species interchangeability (a fundamental requirement of neutrality[5,10]) is inferred at the community or the pair level. Similar evidence from a strikingly different dry forest has been linked to the focused, stable competition of a temporal niche dynamic[11-13]. Taken together with these earlier findings, the results reported here establish a potentially widespread and important role for species interaction in the diversity and maintenance of natural communities that must be considered when inferring process from pattern.**

Linking process and pattern is a central tool of ecology, providing access to ecological questions made otherwise inaccessible by the scale and complexity of natural systems[14,15]. Over the course of the last decade one pattern, the distribution of individuals among species in a community, has carried a debate on putative causal processes that has implications for both ecological and evolutionary inference. Contention has raged over whether community structure and assembly are dominated by 'neutral' processes in which demographic stochasticity is the principal determinant of species coexistence over time, or by niche processes, in particular competition, where interactions between species determine entry, abundance and persistence in communities. Initial claims for neutrality rested upon the capacity for underlying theory to produce a [skewed] lognormal species abundance distribution (SAD)[5]; but either niche or neutral processes may produce the ubiquitous lognormal SAD[7,16-19]. Nor can it be assumed that a lognormal nonetheless may reflect neutrality in nature: under natural conditions, a perfectly good lognormal SAD has been observed supported by significantly non-neutral dynamics[8].

Stochasticity is not a new concept in community ecology, having been incorporated with success into niche-based models as random components in the variation of environmental factors[7,18,19]. In neutral theory stochasticity is assumed to dominate in the demography of the species comprising a community, such that species effectively function as interchangeable units. Neutrality is thus the opposite extreme to niche-defined dynamics in which the differences between species are the basis of species coexistence[20,21]. Determining the relative arenas in nature in which one or the other is more likely must be a central part of any useful integration of the two views into the wider canon of ecology. One suggestion has been that since more similar species would be likely to possess the ecological equivalence defining neutrality, a general class of such taxa should be those species in a community with the maximum degree of shared evolutionary history, co-existing congeners[22]. A pattern of this sort would be of great interest for plant communities, where congeneric pairs and groups make up on average 30% of species[23], although co-occurring congeneric species are certainly not limited to the plant world[22].

A recent study compared abundances of pairs of woody species forming terminal dichotomies in the community-level phylogeny of a Mexican tropical dry forest, reasoning that such species pairs are likely be more similar to one another than either is to any other species in the community[24,25]. Analyses of these data showed that co-existing congeners could not be assumed to be interchangeable, and that the data instead support a conclusion of competition focused within these species pairs[8,13]. Here we investigate the distribution of such fractional abundances of similarly paired tree species in a very different habitat, that of the tropical wet forest of Barro Colorado Island (BCI), and find the same patterns as those revealed in the previous study.

The 50 ha plot on BCI and the sample area at Chamela Biological Station, the site of the México study, differ greatly in physical and biological characters. Both sites are near sea level, but the BCI plot is on a flat-topped island, with the elevational range of the plot itself varying only ~40 m over its entirety. In contrast, the México sample encompasses an elevational range of >140 m. BCI's greater water availability, 2623 mm/yr[26] vs Chamela's 731 mm/yr[27], supports a similarly wider range of stem diameter and tree heights: maximum tree diameter at BCI is ~250 cm while that of Chamela is only ~ 60 cm; the BCI tree data are divided into three growth form categories (understory, mid-sized and canopy tree), while there is no practical need for this with the short-statured forest of Chamela. Although connected by contiguous land mass, the 220-228 tree species of the BCI study and the 190 of the Chamela study have only 7 species in common.

Paralleling the earlier study[8], we calculated for BCI the fractional abundance, $r$, for mature individuals of each pair of species forming a terminal dichotomy in the phylogeny of the community shown in Supplementary Figure 1; $r = n_c/(n_c + n_r)$, where $n_c$ is the number of established (mature) individuals of the more common member of the pair and $n_r$ of the less common. The distribution of $r$, or splitting function, has been shown elsewhere to contain the signature of focused competition when the pairs are chosen from ecological bifurcations[7]. However, if species are equivalent and there is no niche structure or focused interaction, then composing pairs according to any meaningful biological algorithm should yield a distribution of fractional abundance indistinguishable from that of pairs chosen at random from the community. At Chamela it did not; nor does it at BCI.

Species equivalence, interchangeability, is rejected as the explanation of relative abundances of species in both studies. The fractional abundance distribution of the 29 congeneric pairs found in the BCI community level phylogeny differs from the appropriate null model (p = 0.017; Figure 1a), showing abundances within most-closely-related congeneric pairs to be more equitable than for randomly selected species pairs. As in the México study, this effect falls away above the level of genus: the fractional abundance distribution of species forming confamilial terminal dichotomies in the BCI data does not differ from the distribution of randomly selected pairs (p > 0.25; 15 pairs; Figure 1b).

Given the differences between the BCI plot and the Mexican site, we were surprised at how easily the patterns emerged from the BCI data, and how similar the results here are to those of the earlier study[8,13]. Prior to initiating our analyses, we had considered pairing species within tree growth form type in order to examine pattern. Pattern emerged without that refinement, consistent with a sorting algorithm operating before such ecological differences take effect, early in recruitment, an additional factor parallel with the stable competitive processes of temporal niche dynamics linked to the Chamela results[11,13].

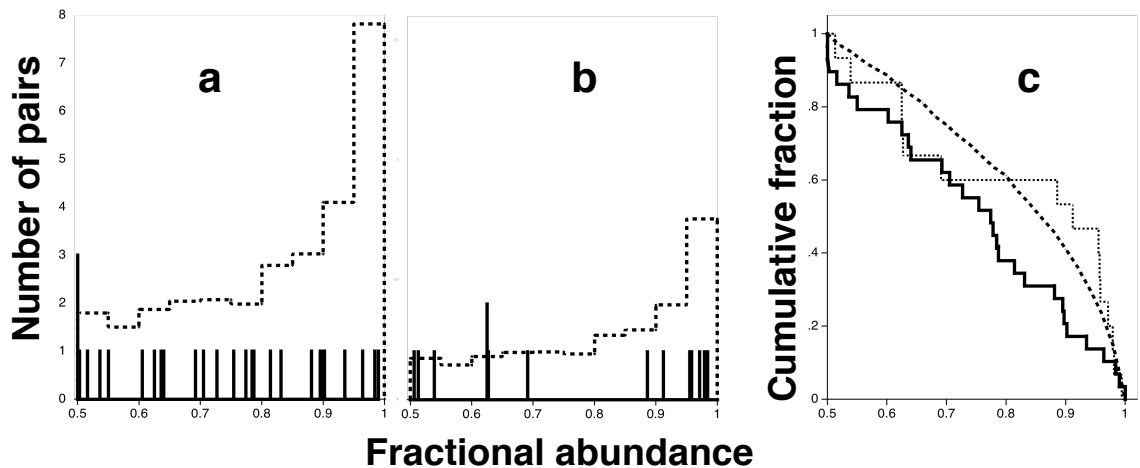

**Figure 1. Fractional abundance distributions.** a. The distribution of fractional abundances of congeneric pairs is significantly different than expected by chance (2-stage $\delta$-corrected Kolmogorov-Smirnov test[28], p = 0.017, N = 29). b. The fractional abundance distribution of non-congeneric pairs does not differ from random (p > 0.25; N = 15). Observed fractional abundance values are shown as idiograms along the x-axes of parts a and b; the dashed line in each shows the null model scaled to the sample size for the comparison. c. Cumulative distributions of null model, congener pairs and non-congeneric pairs used in the statistical analyses. The null model is represented by the heavily dashed curve, fractional abundance distributions for congeneric pairs by the solid black line and non-congeneric fractional abundances by the lighter dashed line. Data for this figure were drawn from the 2005 BCI plot census; 1985, 1990, 1995 and 2000 censuses have slightly different species complements but give similar results, shown in Supplementary Figure 2 and Supplementary Table 1.

The distribution of congener fractional abundances is not consistent with simple similarity within pairs, such as would be expected from interchangeability within the terminal dichotomy [or within the genus as a whole; Supplementary Note 1] (p << 0.01; Figure 2a).

Similarly, congener fractional abundances do not fit a model of competitive exclusion, whether competition is assumed either to be strong (p < 0.008) or weak (p < 0.01; Figure 2b).

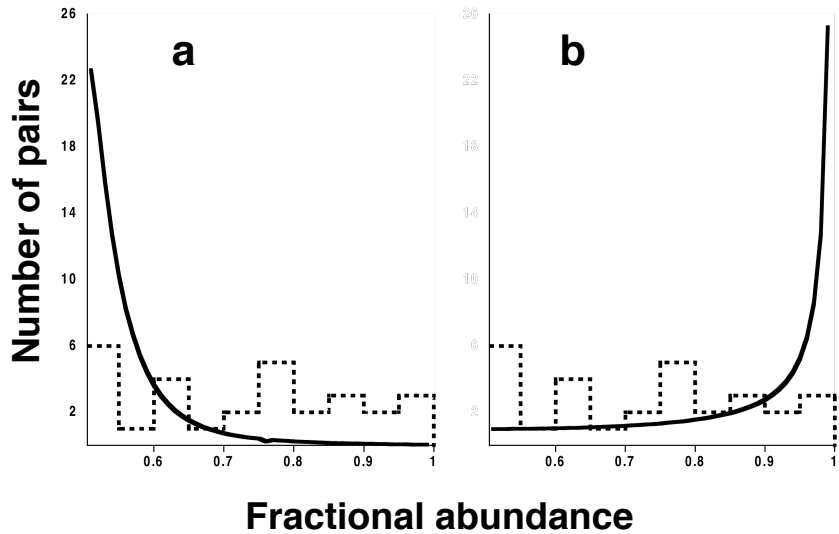

**Figure 2. Alternative hypotheses**.  a. Two-species interchangeability.  The distribution of congener fractional abundances is shown here as a histogram; the solid line represents the distribution expected from two-species interchangeability where a stabilising mechanism is present[10,29].  If the two species were interchangeable, the fractional abundance distribution would peak where population sizes are similar, at 0.5.  The distribution of fractional abundances of most-closely-related congener pairs in the 50 ha plot at Barro Colorado Island is fairly uniformly spread across the range of possible values, differing significantly from that expected from interchangeability within pairs (p << 0.01).  b. Competitive exclusion.  It is similarly unlikely that the observed distribution of congener pairs can be explained by the left-skewed curve (loading most heavily in the 0.9 to 1.0 bin) expected from progressive competitive exclusion (p < 0.01 for weak competition [black line], which gives a less extreme curve than does strong competition).  The areas under the histogram and those under the curves are equivalent.  See Supplementary Note 1 for discussion of the models used here.

Finally, the observed differences cannot be inferred to be a function simply of some general quality of being in a congeneric pair or group.  The species abundance distribution of the congeneric species used in the above analyses does not differ from that of the full complement of species in the BCI community (p > 0.95; Figure 3).  Consistent with this and in parallel with the primary finding reported here, the fractional abundance distribution of congeneric pairs differs from a null model in which pairs are constructed of random draws from a list containing only those same congeneric species.  Pairs selected randomly from a list restricted to only those species occurring in congeneric terminal dichotomies produces a left-skewed curve similar to that drawn from the full community shown in Fig 1, and similarly differs significantly from the observed distribution of congeneric fractional abundances (Kolmogorov-Smirnov two sample test; p < 0.04). The same pattern was found at Chamela[8].

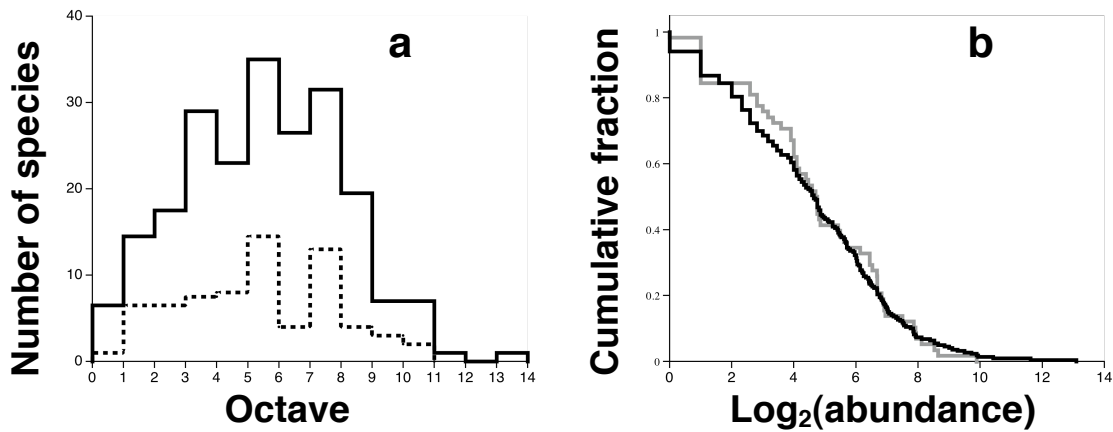

**Figure 3. Species abundance distributions**. a. The upper histogram (solid line) represents abundances of mature individuals of all species within the 50 ha plot. The lower histogram (dashed line) shows abundances of mature individuals of only those species that occur in a terminal dichotomy with a congener partner. The method for determining the exhibited lognormal distributions and the x-axis label 'Octave' follow from Preston[30]; the upper limit of each bin is 2 to the power of the octave value. b. Cumulative distributions of the two SADs. The full community complement is shown by the black line, the congeners by the grey line. Comparisons were of the individual species abundance values for each dataset; n = 220 in the full complement, n = 58 for the congeners. The two distributions cannot be inferred to differ (Kolmogorov-Smirnov; p > 0.95).

In this study we present a second instance of the observation that species paired by phylogenetic similarity have a significant ecological relationship differing from that expected by chance. The patterns revealed indicate interaction between species and consequent niche dynamics allowing coexistence of similar species in this wet forest, as in the earlier study of tropical dry forest[7,8,11,18]. The interaction indicated between most closely related congeners might be supported by either spatial or temporal habitat variation[21]; in the parallel Chamela study, independent evidence implicates the stable competition of temporal dynamics[11,13].

**Methods**

Barro Colorado Island is one of the world's most intensively studied tropical forest communities. Information on the species complements and abundances in the Center for Tropical Forest Sciences 50 ha plot on Barro Colorado Island were retrieved from the archived data found at http://ctfs.si/edu/datasets/bci. The community-level phylogeny was constructed by using all genetic data for all species available on GenBank. Specifics of the methods and the phylogeny schematic are given in Appendix 1. Abundances are of individuals larger than the minimum size of reproduction, which was estimated by R. Foster through applying similar criteria over the full complement of species. Only those species pairs persisting as terminal dichotomies through all five of the censuses were used in comparisons.

**Supplementary Information** accompanies the paper

**Acknowledgements** A second version of this work based on a molecular phylogeny will be available shortly; all analyses have been completed and show the two methods of phylogeny construction to produce the same conclusions.  We thank Katherine Blundell and Stephen Blundell for their help with PERL, T. Pennington for allowing use of his unpublished molecular phylogeny of *Inga* for the literature based tree, and Steve Hubbell for his encouragement and useful comments throughout.  CKK thanks Chamela Biological Station for its hospitality and acknowledges the US National Science Foundation and the National Geographic Society for partial support during the writing of the manuscript.

**Author Contributions** CKK wrote the paper, constructed the phylogenetic tree from published information and collaborated on the statistical analyses; MGB performed the statistical analyses; JNW collaborated on construction of the phylogeny; JBJ constructed the molecular phylogeny that is being used to edit the next version of this ms. All authors discussed the results and commented on the manuscript.

**Author Information** The authors declare no competing financial interests. Correspondence and requests for materials should be addressed to CKK (colleen.kelly@zoology.ox.ac.uk).

**Supplementary Figure 1. Community level phylogeny of the 50 ha Center for Tropical Forest Science plot on Barro Colorado Island, Panamá.** Although not all species were found in all of the 5 complete censuses[1-3], included in the phylogeny are all species of tree with individuals that could be assumed to be reproductive adults (i.e., an established population) and that were found in any one of the censuses. Habit type of tree species are shown here in upper case letters following the species binomial as U = understory tree, M = mid-sized tree and T = canopy level tree. The tree is presented as subunits for better identification of the terminal dichotomy pairs. Authorities are cited in each figure title.

**Supplementary Figure 1a. Phylogeny base[4-7]**

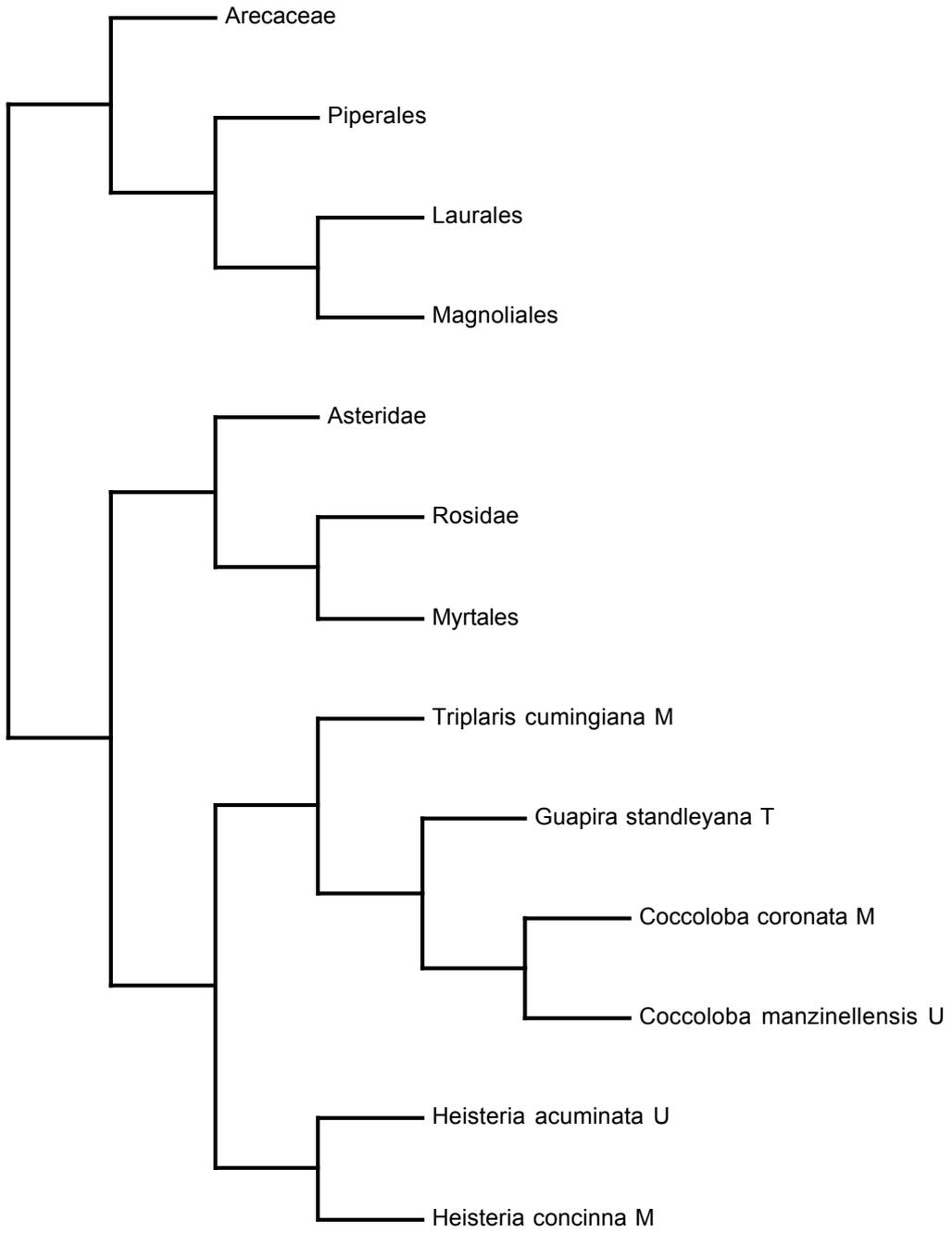

**Supplementary Figure 1b. Arecaceae, Piperales, Laurales and Magnoliales[8-15].**

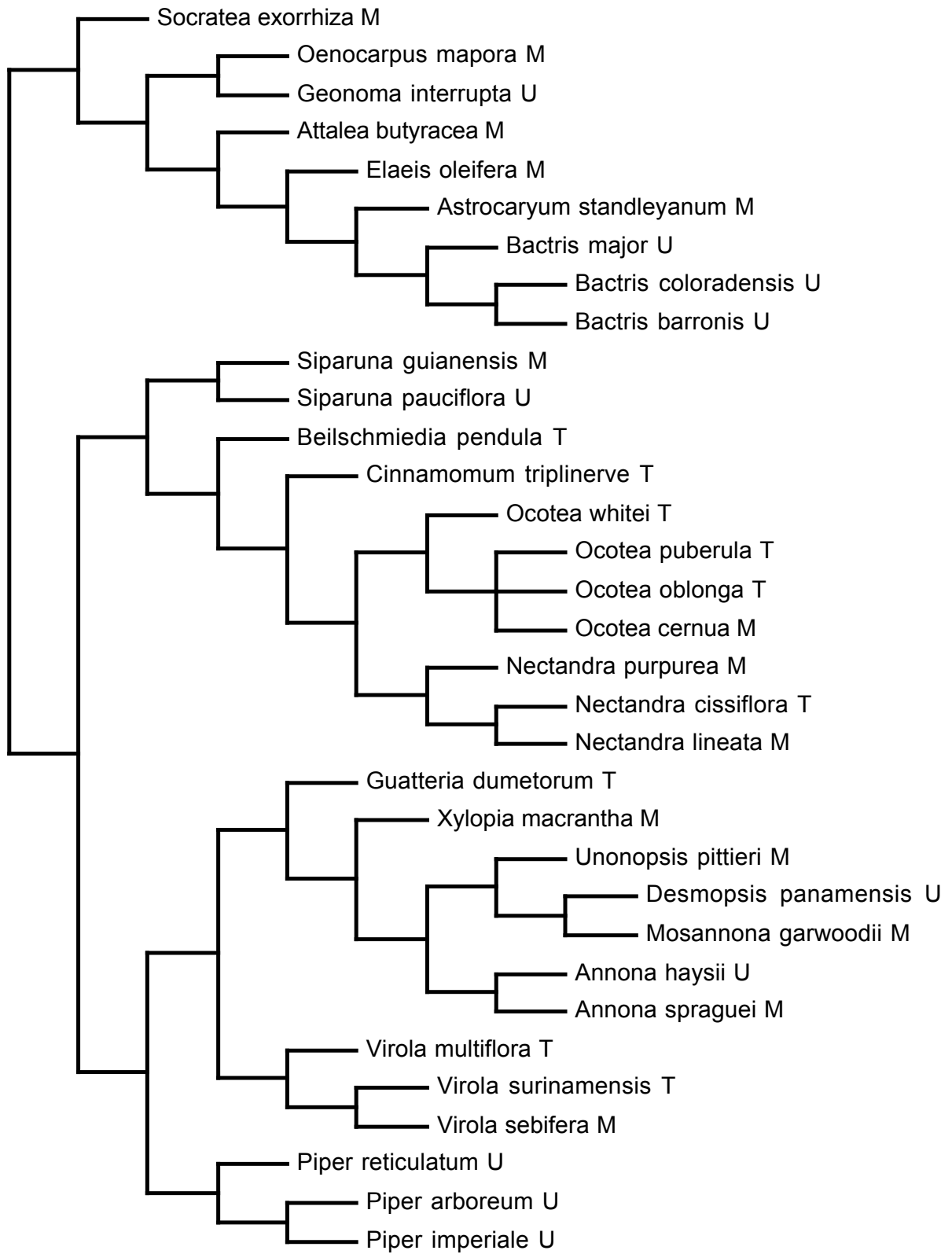

**Supplementary Figure 1c.  Asteridae**[16-21]

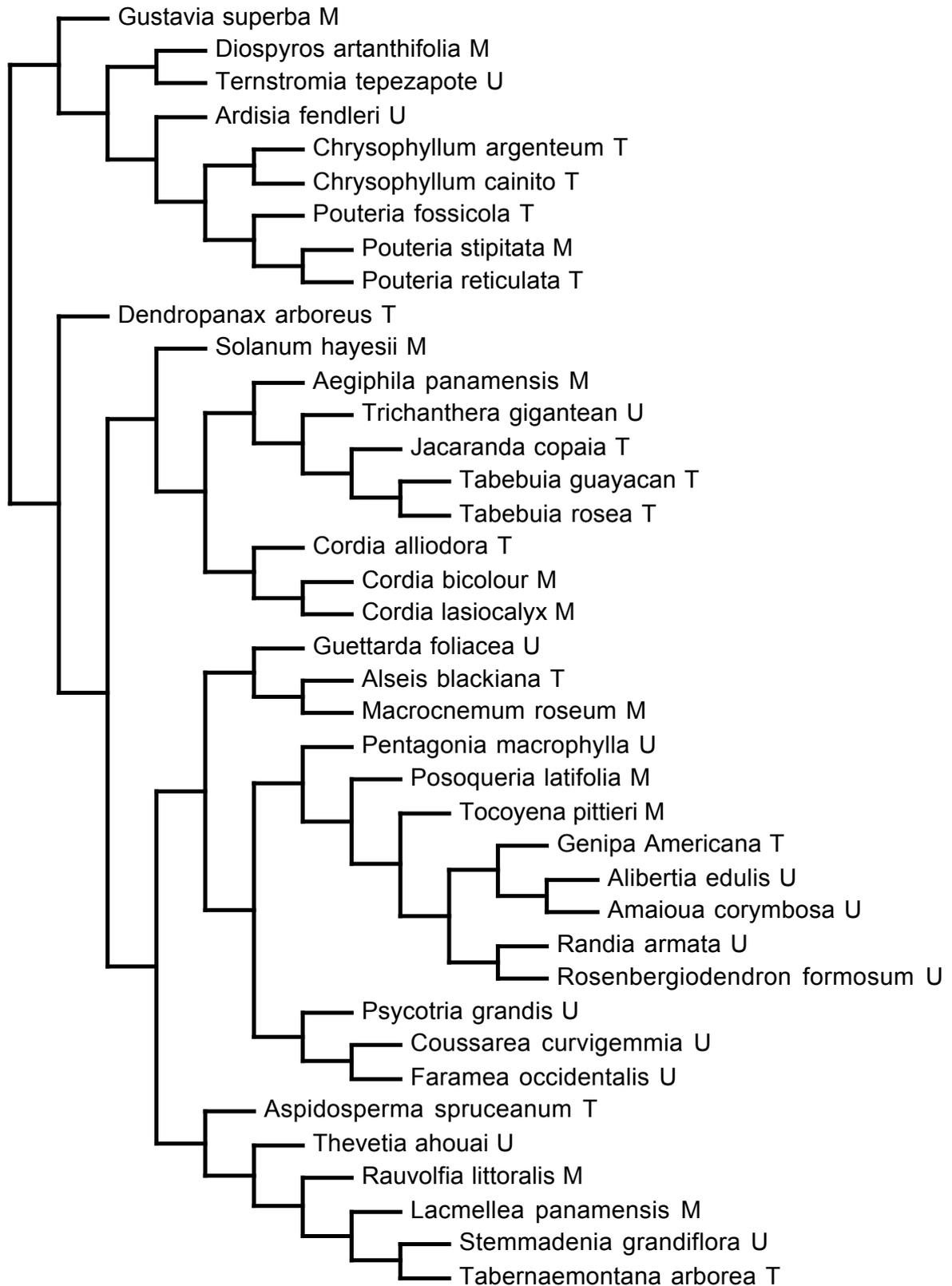

Gustavia superba M
Diospyros artanthifolia M
Ternstromia tepezapote U
Ardisia fendleri U
Chrysophyllum argenteum T
Chrysophyllum cainito T
Pouteria fossicola T
Pouteria stipitata M
Pouteria reticulata T
Dendropanax arboreus T
Solanum hayesii M
Aegiphila panamensis M
Trichanthera gigantean U
Jacaranda copaia T
Tabebuia guayacan T
Tabebuia rosea T
Cordia alliodora T
Cordia bicolour M
Cordia lasiocalyx M
Guettarda foliacea U
Alseis blackiana T
Macrocnemum roseum M
Pentagonia macrophylla U
Posoqueria latifolia M
Tocoyena pittieri M
Genipa Americana T
Alibertia edulis U
Amaioua corymbosa U
Randia armata U
Rosenbergiodendron formosum U
Psycotria grandis U
Coussarea curvigemmia U
Faramea occidentalis U
Aspidosperma spruceanum T
Thevetia ahouai U
Rauvolfia littoralis M
Lacmellea panamensis M
Stemmadenia grandiflora U
Tabernaemontana arborea T

**Supplementary Figure 1d. Myrtales**[5,22,23].

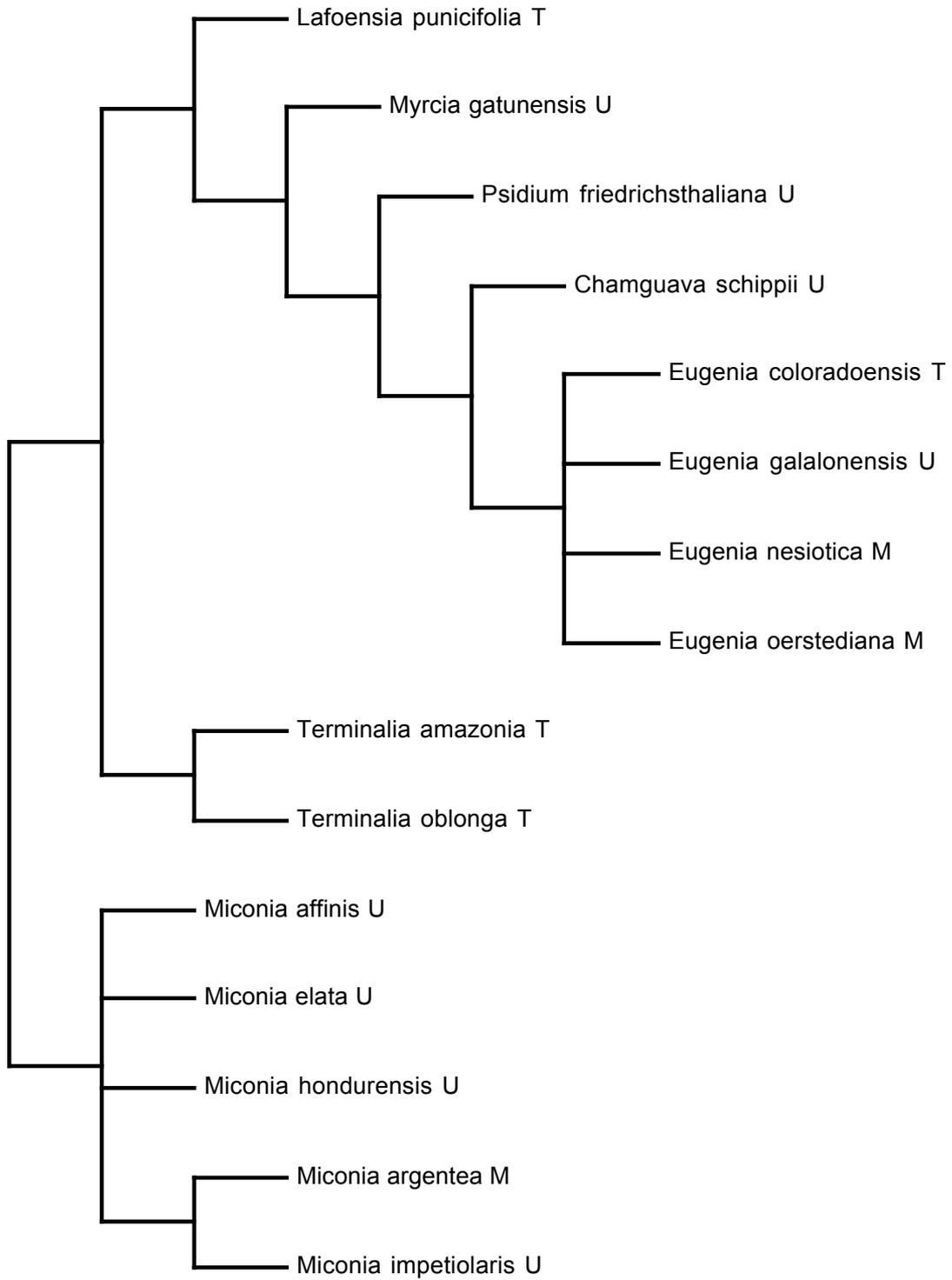

**Supplementary Figure 1e.  Rosales**[24,25].

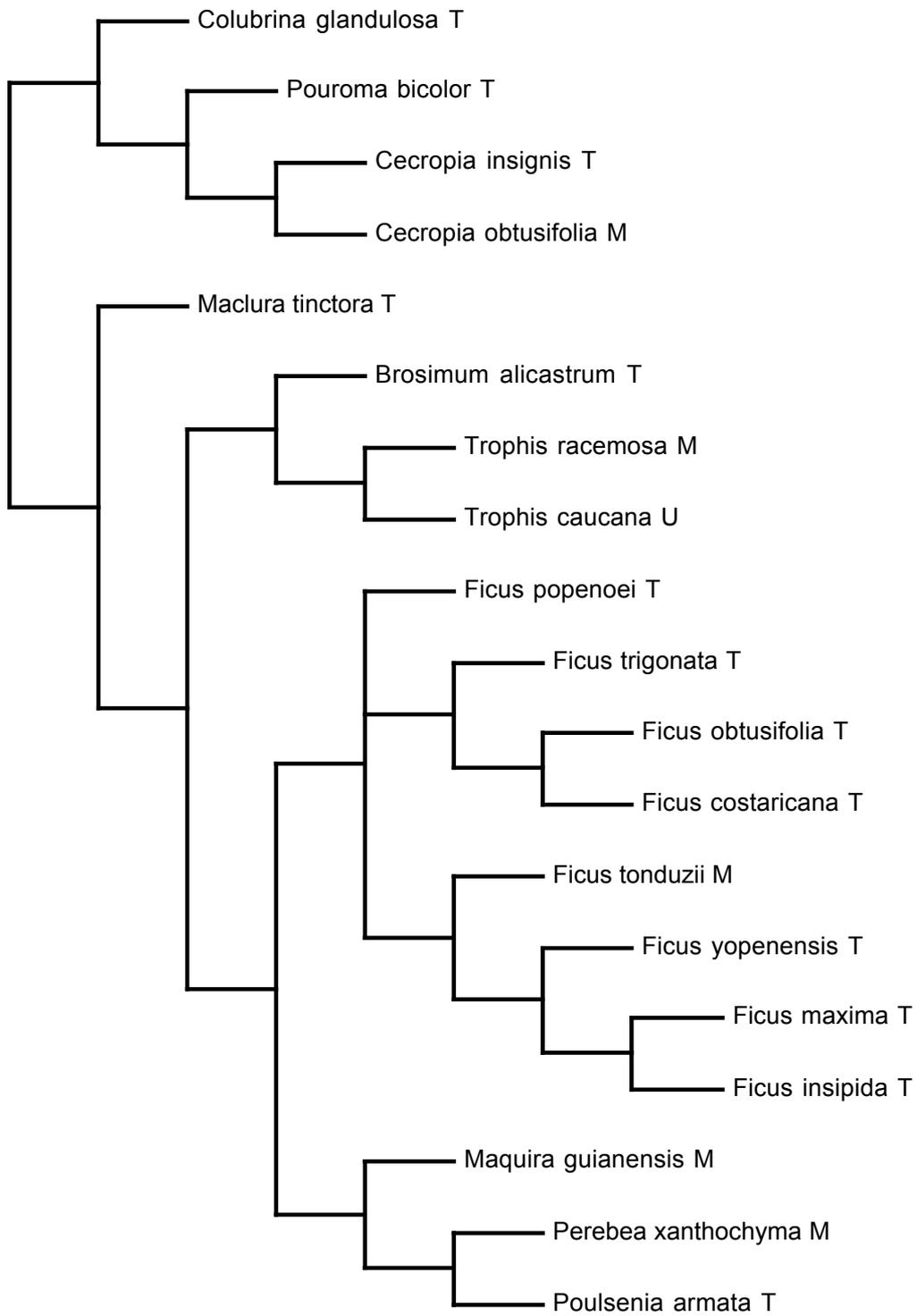

Colubrina glandulosa T
Pouroma bicolor T
Cecropia insignis T
Cecropia obtusifolia M
Maclura tinctora T
Brosimum alicastrum T
Trophis racemosa M
Trophis caucana U
Ficus popenoei T
Ficus trigonata T
Ficus obtusifolia T
Ficus costaricana T
Ficus tonduzii M
Ficus yopenensis T
Ficus maxima T
Ficus insipida T
Maquira guianensis M
Perebea xanthochyma M
Poulsenia armata T

**Supplementary Figure 1f. Fabales**[26-28].

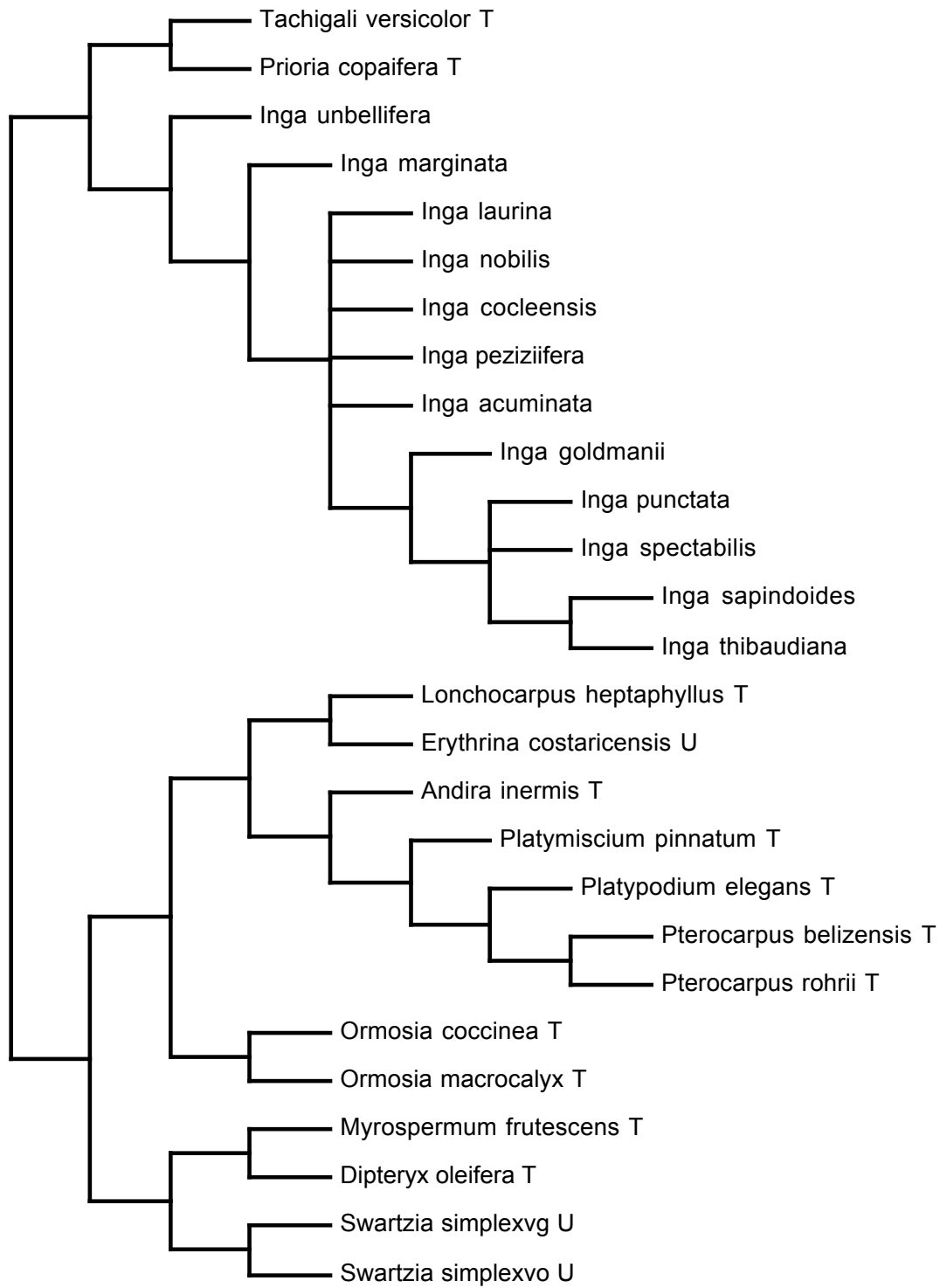

**Supplementary Figure 1g. Malpighiales**[5,29-31]

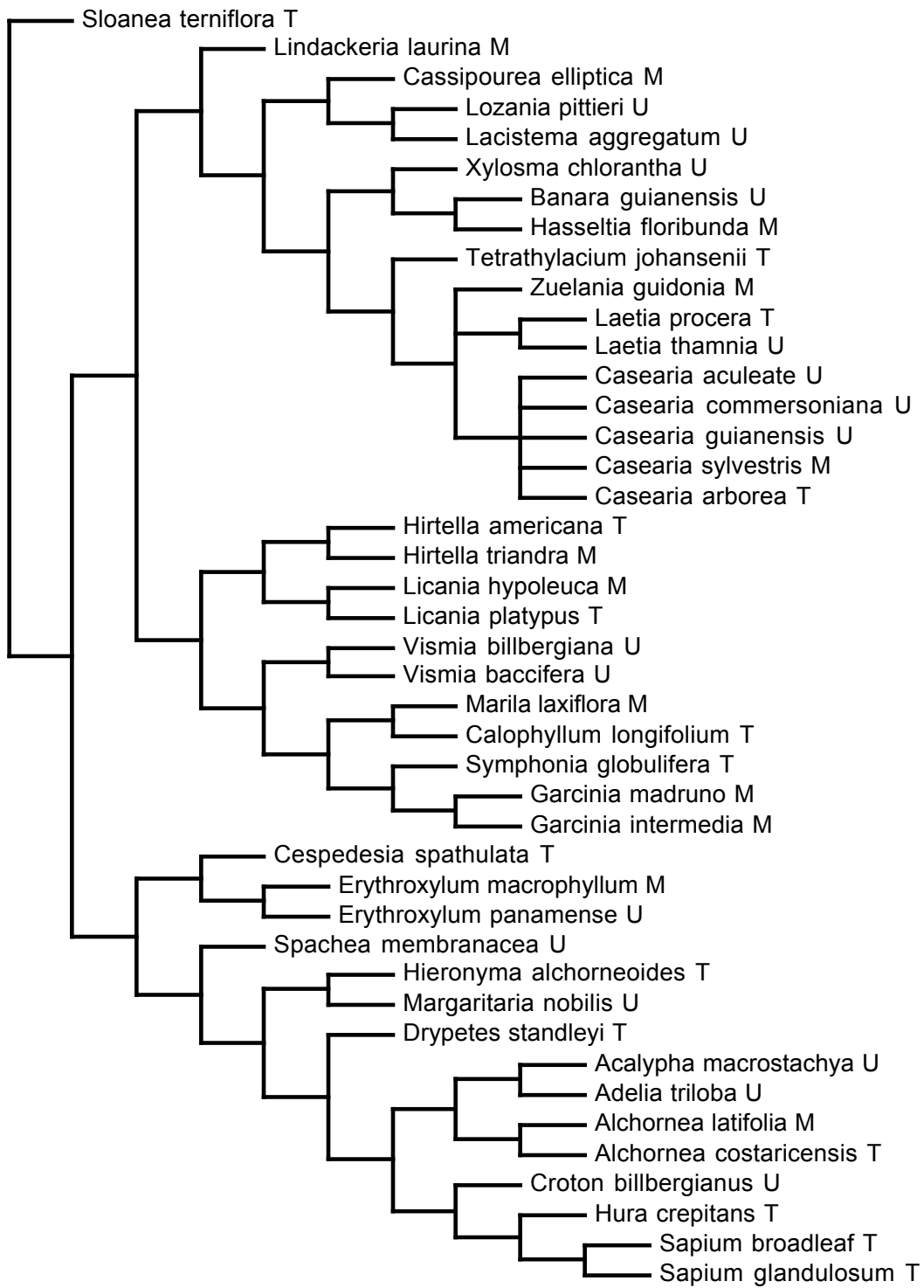

Sloanea terniflora T
Lindackeria laurina M
Cassipourea elliptica M
Lozania pittieri U
Lacistema aggregatum U
Xylosma chlorantha U
Banara guianensis U
Hasseltia floribunda M
Tetrathylacium johansenii T
Zuelania guidonia M
Laetia procera T
Laetia thamnia U
Casearia aculeate U
Casearia commersoniana U
Casearia guianensis U
Casearia sylvestris M
Casearia arborea T
Hirtella americana T
Hirtella triandra M
Licania hypoleuca M
Licania platypus T
Vismia billbergiana U
Vismia baccifera U
Marila laxiflora M
Calophyllum longifolium T
Symphonia globulifera T
Garcinia madruno M
Garcinia intermedia M
Cespedesia spathulata T
Erythroxylum macrophyllum M
Erythroxylum panamense U
Spachea membranacea U
Hieronyma alchorneoides T
Margaritaria nobilis U
Drypetes standleyi T
Acalypha macrostachya U
Adelia triloba U
Alchornea latifolia M
Alchornea costaricensis T
Croton billbergianus U
Hura crepitans T
Sapium broadleaf T
Sapium glandulosum T

**Supplementary Figure 1h. Sapindales and Malvales**[6,32-35].

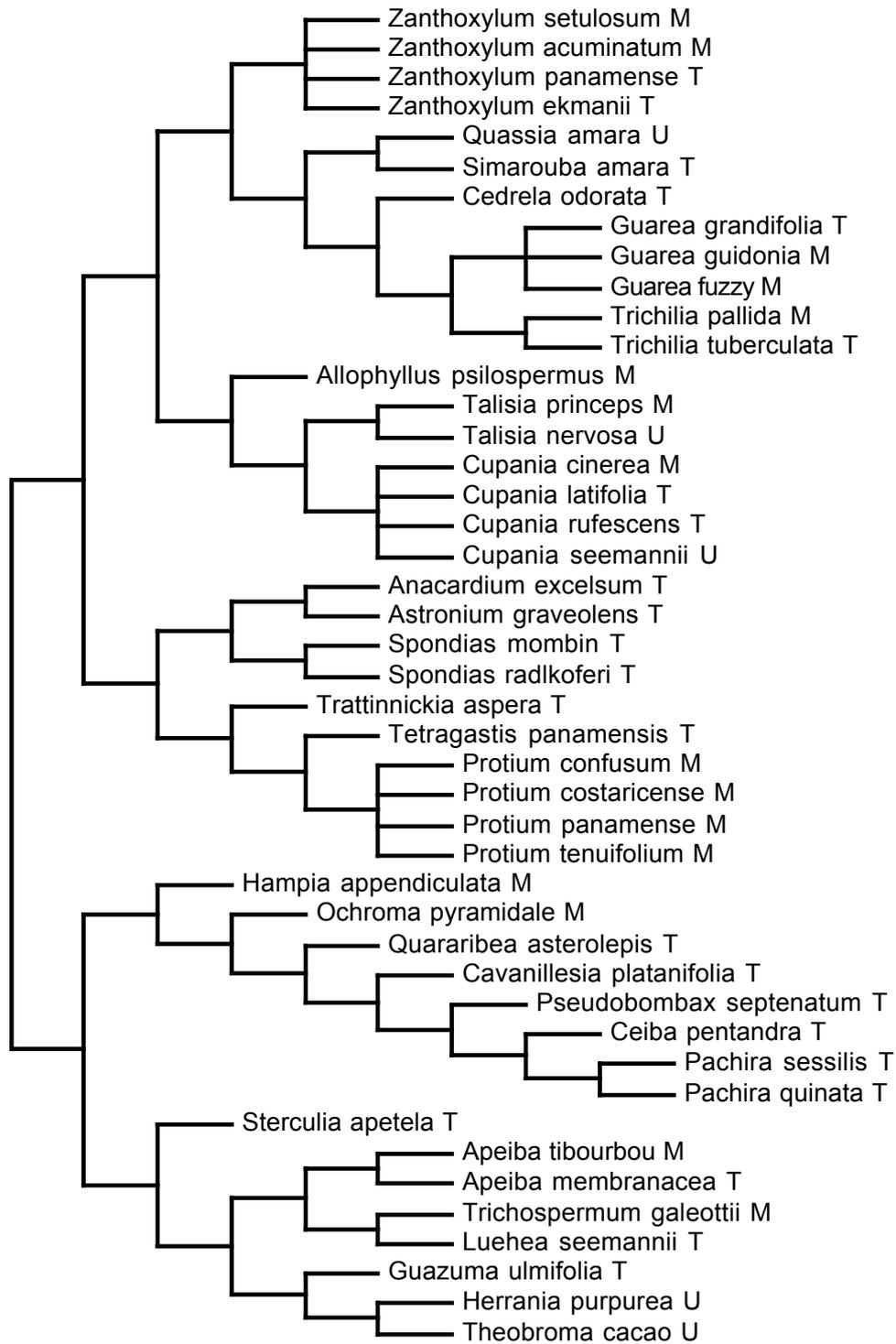

**Supplementary Figure 2.  Fractional abundance distributions.**

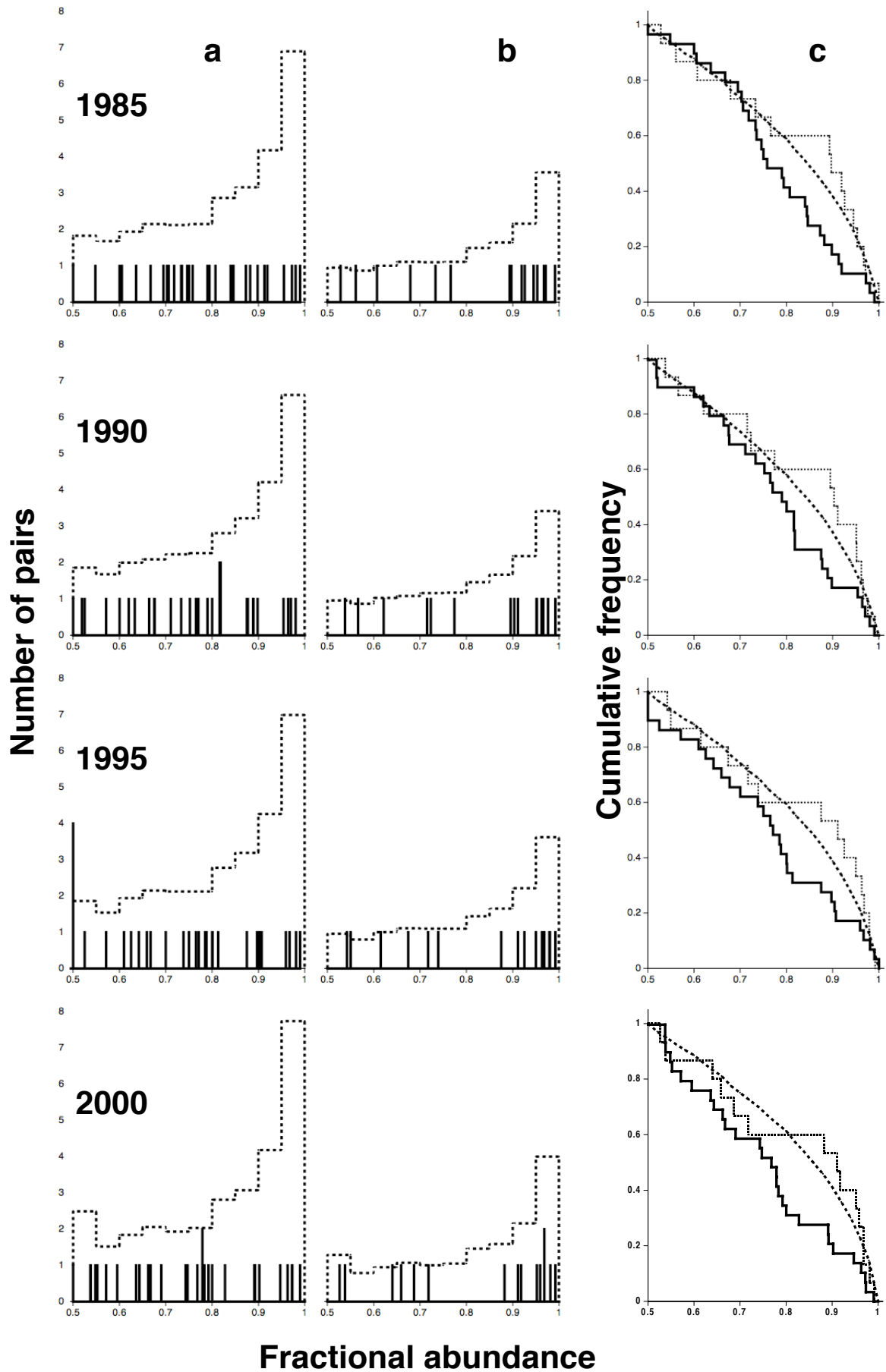

**Supplementary Figure 2. Fractional abundance distributions**. The four fractional abundance distributions shown, together with the 2005 data in the main text, complete the five full censuses that have taken place for the BCI 50 ha plot since 1985. In the figure above, each row gives data from the same census, with the census year to the far left of the row; each column, labelled with a lower case letter at the top of the figure, shows a particular comparison. In all parts, the heavy dashed line represents the null model; the vertical black bars are an idiogram of fractional abundance value of the individual pairs. a. *Fractional abundance values and null model of congeneric pairs*. b. *Fractional abundance values and null model of non-congeneric pairs*. c. *Cumulative distributions of congener pairs (solid line), null model (heavy dashed line) and non-congeneric pairs (lighter dashed line)*. The species complement varied slightly from year to year (between 228 and 220 species), and the null model was drawn from the species complement for the year in question.

**Supplementary Table 1. Kolmogorov-Smirnov comparisons of fractional abundance distributions on a census by census basis.** Comparisons against a null model were performed with a 2-stage, $\delta$-corrected Kolmogorov-Smirnov test; a simple Komolgorov-Smirnov test was used to compare congener vs non-congener fractional abundance[1]. For the relatively small sample sizes available from these data, probability tables go only as high as 0.25. For all five comparisons of congener vs non-congener distributions, the critical difference between the two (maximum separation) would have given a $p < 0.05$ if the sample size of non-congeneric pairs had been as large as for Chamela data, rather than the 15 available from the BCI data.

| Year | Congeneric pairs vs null model | Non-congeneric pairs vs null model | Congneric vs Non-congeneric pairs |
|---|---|---|---|
| 1985 | 0.055 | >> 0.25 | 0.094 |
| 1990 | 0.021 | >> 0.25 | 0.053 |
| 1995 | 0.01 | >> 0.25 | 0.0152 |
| 2000 | 0.007 | >> 0.25 | 0.0152 |
| 2005 | 0.017 | >> 0.25 | 0.090 |

**Supplementary Note 1. Dynamics within congeneric pairs**

If species interchangeability (neutrality[1,2] or species symmetry[3]) is correct, then drawing pairs of species from a guild of tropical trees will yield a fractional abundance distribution identical to drawing pairs at random, whatever the pair selection algorithm may be. Terminal congeneric pairs do not conform to this for either the BCI or Chamela tree communities. Here we discuss the shape of the fractional abundance distribution in the context of general ecological expectations for two ecologically and evolutionarily similar species.

*Interchangebility (Figure 2a)*
To generate an expectation wherein interchangeability functions at the level of congeners, we have supposed that two species (terminal congeneric pairs) are inhabiting the same niche, in the most general sense, and set up two-species Lotka Volterra equations in which the parameters for the two species were identical. Strong stabilising mechanisms have been shown empirically for this site[4] and we incorporated a stabilising mechanism for each species; such stabilising mechanisms operate within a neutral dynamic in which species but not individuals are interchangeable[2]. Under these conditions, the most probable fractional abundance is 0.5. If N pairs have probability 0.5 then the fractional abundance is given by a binomial distribution.

At both Chamela and BCI the majority of genera represented by more than one species are represented by two species, with a number of triplet and very few multiple species groups above that. If the guild is 3 or more congeneric species, then again if there are N equivalent species and a strong stabiliser is built in to permit coexistence, then the most probable allocation of a site to species $n$ is 1/N. The general distribution over sites is a multinomial and again drawing a pair from the guild will have a maximum probability of fractional abundance at 0.5. The distribution may be a bit more complicated than the binomial, but the general form of the expected distribution will still be much like that shown in Figure 2a.

Now suppose we relax the identity (or Chesson symmetry) of the species by letting the carrying capacities (or equivalent parameters) differ by some moderate amount. The effects of all other species in the forest are of course supposed to contribute globally or maybe diffusely to the carrying capacities. If the ratio of carrying capacities for our species is, say, 2:1 the probabilities in the Lotka Volterra equations will still converge and they will now converge on 66/33 and a fractional abundance of 0.66. The binomial distribution will now be centred on 0.66 and fall away on both sides – more like the sort of thing we see and we could not rule it out.

These species are not identical, but would we call them similar? We might well – the populations are interdigitated and certainly if the carrying capacities were only 1% different we would. In a sense the argument becomes circular – the species might be called similar if the fractional abundance distribution is for probabilities and hence carrying capacities much closer than for a randomly drawn pair. We have of course applied different criteria to determine similarity so we can claim that similar species have similar carrying capacities … we do not reject the suggestion that similarity is important – we became interested in congener pairs in the first place to exploit the capacity of close relatedness in factoring out a lot of inessential aspects of species coexistence[5-7]. But regardless of all else, the real question remains: these species are similar but not identical and yet they coexist – How?

*Transient coexistence (a version of competitive exclusion) (Figure 2b)*
Suppose species *a* is occupying a certain large number of sites *N*. A similar congeneric species is added and outcompetes it. Every time an individual of species *a* dies it has a high probability of being replaced by an individual of species *b*. The probability of any individual of species *a* dying is the same small constant each year – the population of *a* dies away exponentially. Then early on the proportion of *a* is close to 1, but this does not last long – only one half life in this simple model – after which the proportion of *b* is greater and this lasts almost forever (forever with an infinite population or a continuous distribution). If for any congeneric pair we measure at a randomly sampled time we are very likely to find the winner present at a very high fractional abundance.

The shape is easily calculated. As a function of time the fractional abundance is $e^{-kt}$ up to one half life and thereafter $1 - e^{-kt}$. If we cannot pick out two species after *n* half lives, then the distribution of fractional abundances *r* randomly sampled is

$$1/(1-r) + (1/n)(1/r - 1/  (1-r))$$

A variant in which the proportion of dead species *a* replaced by *a* depends on the density of species *a* corresponds to weaker competitive advantage for the invader, species *b*. This makes only a qualitative difference; the symmetric function $\dfrac{1}{r(1-r)}$ for this case is plotted in Figure 2b in the body of the paper.

This exponential behaviour is a very simple, yet plausible. (We note that the resulting fractional abundance distribution is like that of the non-congeneric pairs or that of pairs drawn at random from the data. We would not wish to imply that this similarity of shape identifies the mechanism.) Sampling competing pairs along the road to competitive exclusion is likely to yield fractional abundances at the high end (0.9 – 1 with sufficient data) and the more equitable fractions we see in the real data are not likely.